\documentclass[preprint,aps,amsmath,superscriptaddress,nofootinbib,tightenlines,floatfix]{revtex4}
\usepackage{graphicx}
\usepackage{bm}
\usepackage{epsfig}

\newif\ifpdf
\ifx\pdfoutput\undefined
\pdffalse 
\else
\pdfoutput=1 
\pdftrue
\fi

\def\bnslash{\bar n\!\!\!\slash}

\def\OMIT#1{}

\newcommand{\CH}[2]{\chi_{#1,#2}}

\newcommand{\bCH}[2]{\overline\chi_{#1,#2}}

\newcommand{\nn}{\nonumber} 

\newcommand{\bn}{{\bar n}}
\newcommand{\bea}{\begin{eqnarray}}
\newcommand{\eea}{\end{eqnarray}}

\newcommand{\bnP}{\bar {\cal P}}
\newcommand{\ppP}{{\cal P}_\perp}
\newcommand{\bnPd}{\bar {\cal P}^{\raisebox{0.8mm}{\scriptsize$\dagger$}} }
\newcommand{\cP}{{\cal P}}

\newcommand{\LQCD}{{\Lambda_{\rm QCD}}}

\newcommand{\SCETa}{SCET$_{\rm I}$}

\begin{document}

\ifpdf
\DeclareGraphicsExtensions{.pdf, .jpg}
\else
\DeclareGraphicsExtensions{.eps, .jpg,.ps}
\fi


\preprint{\vbox{ \hbox{CMU-HEP-04-03}  }}

\title{Flavor-singlet light-cone amplitudes and radiative $\Upsilon$ decays in SCET} 

\author{Sean Fleming}
\affiliation{Department of Physics, Carnegie Mellon University,
      	Pittsburgh, PA 15213\footnote{Electronic address: spf@andrew.cmu.edu}
	\vspace{0.1cm}}
	
\author{Adam K. Leibovich}
\affiliation{Department of Physics and Astronomy, 
	University of Pittsburgh,
        Pittsburgh, PA 15260\footnote{Electronic address: akl2@pitt.edu}
	\vspace{0.2cm}}

\date{\today\\ \vspace{1cm} }



\begin{abstract}

We study the evolution of flavor-singlet, light-cone amplitudes in the soft-collinear
effective theory (SCET), and reproduce results previously obtained by a different approach. 
We apply our calculation to the color-singlet contribution to the photon endpoint in radiative 
$\Upsilon$ decay. In a previous paper, we studied the color-singlet contributions to the endpoint, 
but neglected operator mixing, arguing that it should be a numerically small effect.
Nevertheless the mixing needs to be included in a consistent calculation, and we do just that in 
this work. We find that the effects of mixing are indeed numerically small. This result combined
with previous work on the color-octet contribution and the photon fragmentation contribution
provides a consistent theoretical treatment of the photon spectrum in $\Upsilon \to \gamma X$.

\end{abstract}

\maketitle

\newpage
\section{Introduction}

The soft-collinear effective theory (SCET)  
\cite{Bauer:2001ew,Bauer:2001yr,Bauer:2001ct,Bauer:2001yt} is a systematic 
treatment of the high energy limit of QCD in the framework of effective field theory. Prior to 
the introduction of SCET this limit of QCD was subject to intense study using various other approaches 
including all order perturbative methods \cite{AdvSeries}. Some of these classic calculations have
been revisited in SCET and their results reproduced~\cite{Bauer:2002nz, Manohar:2003vb}. The
effective theory approach, however, goes beyond the approximations upon which many 
of the previous calculations rely. In particular it is straight forward to include power corrections 
to any process, as was demonstrated in the context of color-suppressed $B$ meson decays
\cite{Mantry:2003uz}, which receive their first contribution at subleading order. In addition SCET
naturally includes nonperturbative effects in the form of matrix elements of operators. This for 
example, gives a consistent explanation for the origin of the shape function in semi-inclusive
$B$ meson decay. In this article we study the radiative decay of the Upsilon, and revisit another 
classic result: namely the evolution equation for light-cone wave functions, also known as the
Brodsky-Lepage equation.

At first sight it may seem strange to be discussing a heavy quarkonium system in the context of a
high-energy effective theory. It is, however, the final state of radiative Upsilon decay which can, in a certain region of phase space, be described by SCET. To describe the $\Upsilon$ system, which is a boundstate of a heavy $b$ quark and $\bar{b}$ quark, we need to consider a different limit of QCD:  the non-relativistic limit. This is sensible because the large mass of the $b$ quark ensures that the 
the typical relative velocity $v$ of the $b$ and $\bar{b}$ in the $\Upsilon$ is small $v \sim 0.3$ allowing for a non-relativistic expansion. Furthermore the production and decay a $b\bar b$ pair can be calculated perturbatively.  In the earliest works on quarkonium, the $v\to 0$ limit was always taken, allowing all the non-perturbative dynamics to be isolated into the wavefunction at the origin.  This approach is now called the color-singlet model, since in an effective theory picture it corresponds to keeping only those operators that create/annihilate the $b\bar b$ in a color-singlet configuration. With the advent of a non-relativistic effective theory of QCD (NRQCD) \cite{bbl,lmr}, this simple picture is replaced by a systematic expansion in operators that scale as higher and higher powers of $v$, where some of the time the quarkonium state can be produced/annihilated in a color-octet configuration.  

The theoretical picture of radiative Upsilon decay that emerges from these considerations is quite
rich. Over some of phase space the decay is described by the annihilation of a $b\bar b$ pair in a color-singlet ${}^3S_1$ configuration into a photon and a pair of gluons with invariant mass on the order of the Upsilon mass. This is well described by an operator product expansion based on NRQCD. However, the situation is more complicated as the photon energy reaches its maximum. In this region of phase space the pair of gluons form a collinear jet back-to-back with the photon, and there arises a possibly large contribution from the annihilation of the $b\bar b$ in a color-octet configuration into a photon back-to-back with a single gluon. Since the decay products in this  ``endpoint" region  are jet-like (i.e. their energy is large relative to their invariant mass) the appropriate effective theory to describe the dynamics of the decay products is  SCET. The $\Upsilon$ is still described by NRQCD.   

The radiative decay of the Upsilon is of particular interest since it  allows for a measurement of the strong coupling constant $\alpha_s$ \cite{Albrecht:1987hz,Bizzeti:1991ze,Nemati:1996xy,Gremm:1997dq}. Furthermore the differential decay rate as a function of the energy fraction $z = 2 E_\gamma/M_\Upsilon$ has been measured, and in each case found to be softer than the QCD predictions. In a series of recent papers~\cite{Bauer:2001rh,Fleming:2002rv,GarciaiTormo:2004jw} this decay has been studied using SCET to describe the endpoint region of the decay rate. First in Ref.~\cite{Bauer:2001rh}, the large Sudakov logarithms for the color-octet channels were resummed for the first time using SCET.  In  subsequent papers \cite{Fleming:2002rv}, we analyzed the color-singet decay in the endpoint region.  This was calculated by Photiadis \cite{Photiadis:1985hn}. However, in  Ref.~\cite{Fleming:2002rv} we ignored the possibility of a jet of a light quark and anti-quark in the final state, and we reproduced Photiadis' results in this limit.  The $\bar q q$ final state has a zero tree-level matching coefficient in the effective theory for this process, but it can be generated by mixing with the gluon jet, and so it must be included in a consistent calculation.  

The main result of this work is the derivation within the SCET framework of the evolution equation for matrix elements of collinear operators that describe the gluon and quark jet final states in radiative Upsilon decay near the endpoint. As a consequence of the factorization of soft physics from collinear physics the evolution of these matrix elements of helicity-zero, flavor- and color-singlet collinear operators is quite general, and should hold for any collinear final state produced from the vacuum. This was pointed out by Photiadis. Thus, the evolution equation for the matrix elements we are concerned with should be the similar to that of the pion lightcone wave function which was first considered in Ref.~\cite{Lepage:1979zb,Efremov:1978rn}. However, the full mixing is not incorporated in those works, since the pion is a flavor non-singlet.  The flavor singlet case was done by Chase~\cite{Chase:hj} in the context of quark-antiquark and gluon-gluon jet production in photon-photon collisions. We reproduce those results using SCET. In Sec.~II we quickly review soft-collinear effective theory, in Sec.~III we introduce the collinear operators that arise in radiative Upsilon decay, in Sec.~IV we calculate the renormalization group equation that governs the running of these operators, in Sec.~V we use the results of the previous section to give the resummed rate for radiative $\Upsilon$ decay, and in Sec.~VI we conclude.

\section{Review of SCET}

We begin with a short review of the parts of SCET that are relevant to this calculation.  In particular, we are only concerned with \SCETa~\cite{Bauer:2002aj}, which describes the interactions of collinear and ultra-soft (usoft) degrees of freedom. In this theory collinear particles have momenta whose lightcone components scale as $p = (p^+, p^-, p_\perp) \sim Q(\lambda^2, 1, \lambda)$, where $Q$ is a large energy scale, and  $\lambda  \ll 1$ is a small expansion parameter. In ${\rm SCET}_{\rm I}$ $\lambda \sim \sqrt{ \LQCD/Q}$ so that the typical invariant mass of an ${\rm SCET}_{\rm I}$ collinear particle is $p^2 \sim Q \LQCD$. Usoft particles have momenta which scale as $k = (k^+, k^-, k_\perp) \sim Q(\lambda^2, \lambda^2, \lambda^2)$, so that the typical invariant mass of a usoft particle is $k^2 \sim \LQCD^2$.  The usoft degrees of freedom interact with the collinear particles without taking the collinear particles off-shell by more than $\sim Q \LQCD$. Furthermore it is only the plus component of the collinear momentum that a usoft particle can change.

Here we are interested in the differential decay  rate for $\Upsilon \to \gamma + X$ as a function of the photon energy restricted to the region where $2 E_\gamma \sim M_\Upsilon - {\cal O}(\LQCD )$. In this regime the final state hadrons have a lightcone momentum componenet of order $M_\Upsilon$, and invariant mass of order $ \sqrt{M_\Upsilon (M_\Upsilon - 2 E_\gamma)}$. Clearly the jet can be described with ${\rm SCET}_{\rm I}$, where $Q = M_\Upsilon$ and the power-counting parameter is  $\lambda= \sqrt{1 - 2 E_\gamma/M_\Upsilon}$. The $\Upsilon$ particle can be treated in NRQCD~\cite{bbl,lmr}, where large fluctuation about the heavy quark mass are integrated out, leaving only modes with momentum of order $mv^2 \sim \LQCD$, where $v\sim 0.3$. These usoft modes can interact with both the heavy quarks in the inital state and the collinear particles in final state.

By matching QCD onto SCET the large scale $Q$ is integrated out.  In
practice, the matching procedure is to calculate matrix elements in
QCD, expand them in powers of $\lambda$, and match onto products of
Wilson coefficients and operators in SCET. Thus it is important to be
able to deduce the SCET operators which can arise at a given order in
$\lambda$.  Field theory generally allows all operators that are
consistent with the symmetries of the theory.  As explained in detail
in Ref.~\cite{Bauer:2001yt}, the symmetry of SCET which restricts the
operators that can arise is gauge invariance. Specifically, SCET is invariant under
two types of gauge transformations: collinear and usoft. Under collinear gauge 
transformations the usoft fields remain invariant, while the collinear fields transform 
in the usual manner. Under usoft gauge transformations the usoft fields transform 
in the usual manner, and the collinear fields undergo a global color rotation. 

The collinear fields in SCET are the fermion field $\xi_{n,p}$, and the gluon field 
$A_{n,q}^\mu$. These fields are labeled by the lightcone direction
$n^\mu$, and the large components of the lightcone momentum ($\bn\cdot
q, q_\perp$).  The  femion field contains a term $\xi^{+}_{n,p}$ that annihilates particles,
and a term $\xi^{-}_{n,-p}$ that creates antiparticles. For the construction of gauge invariant
operators we will find it convenient to  make use of the SCET  collinear Wilson line,
\begin{equation}
W_n(x) = \bigg[ \sum_{\rm perms} {\rm exp} 
  \left( -g_s \frac{1}{\bnP} \bn \cdot A_{n,q}(x) \right) \bigg] \,.
\end{equation}
The operator ${\cal P}^\mu$ projects out the momentum label~\cite{Bauer:2001ct} of fields
that sit to the right of the operator. We will use the convention that ${\cal P}^\mu$ only acts
on those fields in the square brackets. Generally 
$\big[ f(\bnP) 
\phi^\dagger_{q_1} \cdots  \phi^\dagger_{q_m} \phi_{p_1} \cdots  \phi_{p_m} \big] = 
f(\bn\cdot p_1 + \cdots+ \bn\cdot p_n - \bn\cdot q_1-\cdots-\bn\cdot q_m)\phi^\dagger_{q_1} \cdots \phi_{p_m} $ ,
where $\bnP\equiv\bar n \!\cdot\! {\cal P} $.
The conjugate operator $\bnPd$ acts on fields that sit to the left of the operator, and 
projects out the sum of labels on conjugate fields minus the sum of labels on fields.
In the usoft sector there is a usoft fermion field $q_{us}$,  and a usoft gluon field
$A^\mu_{us}$. 

Operators in SCET are
constructed such that they are gauge invariant under both collinear  and usoft 
gauge transformations. For example, under collinear-gauge transformations 
$\xi_{n,p} \to U_n \xi_{n,p}$ and $W_n \to U_n W_n$, so the combination
\begin{equation}
W^\dagger_n \xi_{n,p}
\end{equation}
is collinear-gauge invariant. Furthermore it is convenient to introduce a 
delta function which fixes the labels of the combination of fields above:
\begin{equation}
\CH n \omega  \equiv  [\delta_{\omega, \bnP} W^\dagger_n \xi_{n,p}], 
\end{equation}
where it is understood that we will include a sum over $\omega$ for 
each $\chi_{n,\omega}$ in an operator. The Wilson coefficient will in 
general depend on the label momentum $\omega$, which will result in a 
convolution of the short distance coefficient with the SCET operator.
The combination above still transforms under a usoft-gauge transformation 
$\CH n \omega \to V(x) \CH n \omega $. 

The collinear-gauge invariant field strength is
\begin{equation}
G^{\mu\nu}_n \equiv -\frac{i}{g_s} \bigg[ W^\dagger \big( i{\cal D}_n^\mu 
  + g_sA_{n,q}^\mu, i{\cal D}_n^\nu+g_sA_{n,q'}^\nu \big) W \bigg] \,,
\end{equation}
where 
\begin{equation}
i{\cal D}_n^\mu = \frac{n^\mu}2 \bnP + \ppP^\mu + 
\frac{\bn^\mu}2 i n\cdot D,
\end{equation}
and $iD^\mu = i \partial^\mu+g_sA^\mu_s$ is the usoft covariant
derivative.  Note that $G^{\mu\nu}_n$ is not homogeneous in the power
counting. The leading piece scales like $\lambda$, and is given by
$[\bnP B^\mu_\perp] \equiv\bn_\nu G^{\nu\mu}_n$, where the perp subscript
on $B$ indicates that the $\mu$ index only has support over
perpendicular components. Simplifying and including a label fixing delta function,
we obtain
\begin{equation}\label{bfield}
B^\mu_{\perp \omega} =  \frac{-i}{g_s} \bigg[ \delta_{ \omega , \bnP }
W^\dagger (\ppP^\mu + g_s (A^\mu_{n,q})_\perp)W \bigg] \,. 
\end{equation}
Under usoft gauge transformations 
$B^\mu_{\perp \omega }  \to V(x) B^\mu_{ \perp\omega } V^\dagger (x)$. 
We use these objects to build the operators we need to match onto SCET
at the endpoint of the $\Upsilon\to X\gamma$ spectrum. For further
examples the reader is referred to Ref.~\cite{Bauer:2002nz}.

\section{SCET operators}

We begin by matching the QCD final states onto SCET operators.  Since we are interested in the color- and flavor-singlet, helicity-zero operators, we have the SCET operators
\bea
\label{leadingops}
O_{g}(\omega_1,\omega_2) &=& \bnP
{\rm Tr}[B_{\perp \omega_1}^\alpha \, B_{ \perp \omega_2 }^\beta] g^\perp_{\alpha\beta} \,,
\nn\\
O_{q}(\omega_1,\omega_2) &=& \bCH n {\omega_1} \frac{\bnslash}{2}\chi_{n,\omega_2},
\eea
where $g_\perp^{\alpha\beta} = g^{\alpha\beta} - (n^\alpha\bar n^\beta + n^\beta\bar n^\alpha)/2$. 
We have introduced an additional factor of $\bnP$ into the gluon operator so that both of the above
operators have the same energy dimension. In addition, both operators scale as $\lambda^2$ in the SCET power counting. They are the complete set of leading color-singlet operators. Each of the operators are convoluted with a short distance coefficient
$\Gamma_{g/q}(\omega_1 , \omega_2 )$, which is determined by matching onto QCD. 

Matrix elements of the operators in Eq.~(\ref{leadingops}) are non-perturbative functions
of the labels $\omega_1$ and $\omega_2$. Consider the matrix element of a collinear 
final state ${\cal F}_{n,p}$ and collinear initial state ${\cal I}_{n,p'}$:
\begin{equation}
\langle {\cal F}_{n,p} | \bCH n {\omega_1} \frac{\bnslash}{2} 
\CH n {\omega_2} | {\cal I}_{n,p'} \rangle \,.
\end{equation}
This can be simplified by introducing $\omega_\pm = \omega_1\pm \omega_2$ and 
$p_\pm = p\pm p'$, and using momentum conservation
\bea\label{lca}
\langle {\cal F}_{n,p} | \bCH n {\omega_1} \frac{\bnslash}{2} 
\CH n {\omega_2} | {\cal I}_{n,p'} \rangle  &=& \delta_{\omega_- , \bn\cdot p_- } 
\langle {\cal F}_{n,p} | \bar\xi_{ n, \omega_1} W  \frac{\bnslash}{2} \delta_{\omega_+  ,  \cP_+}
W^\dagger \xi_{n,\omega_2} | {\cal I}_{n,p'} \rangle
\nn \\
& & \Longrightarrow   \delta_{\omega_- , \bn\cdot p_- } 
{\cal K}_{{\cal FI}} \phi_{\cal F I}(x_+)\,,
\eea
where $\phi_{\cal F I}$ is the light-cone amplitude (LCA) for the transition ${\cal I} \to {\cal F}$, and
is by definition dimensionless. This last requirement on $\phi_{\cal F I}$ forces us to introduce the  constant ${\cal K}_{{\cal FI}}$ which is process dependent and possibly dimensionful. In Upsilon decay ${\cal K}_J = M^2$. To arrive at the last line of the above equation we extend the sum over discrete $\omega_+$ to an integral over continuous  $\omega_+$ and define $x_+ = \omega_+ / \bn\cdot p_-$. As a result all sums over $\omega_+$ are converted to integrals over $x$. In Appendix~\ref{appRPI} we show how this is done using type (a) RPI as defined in Refs.~\cite{Chay:2002vy,Manohar:2002fd}. 

Two specific choices of initial and final state
are familiar. If we choose the incoming and outgoing state to be a proton with momentum $p$,
then $\phi_{\cal F I}$ is related to the parton distribution functions. If, however,
the incoming state is the vacuum, and the outgoing state is a meson with momentum $p$, then  $\phi_{\cal F I}$ is related to the light-cone wave function of the meson. 

In the case of $\Upsilon  \to \gamma + X$ in the large photon energy regime the QCD 
amplitude for $b\bar{b}({\bf 1},{}^3S_1) \to \gamma g g$ matches 
onto a convolution of a short-distance Wilson coefficient and an SCET current~\cite{Fleming:2002rv}
\begin{equation}\label{matching}
J^\mu(z) = \sum_\omega e^{-i(Mv + \bnP n/2)\cdot z} \Gamma^{({\bf 1},{}^3S_1)}_g (\omega; \mu)
J^\mu_{({\bf 1},{}^3S_1)} (\omega ; \mu ) \,,
\end{equation}
where 
\begin{equation}
J^\mu_{({\bf 1},{}^3S_1)} (\omega ; \mu ) = \chi^\dagger_{- \textbf{p}}
\Lambda\!\cdot\! \mathbf{\sigma}^\mu  \psi_{\textbf{p}} 
\textrm{Tr}\big[ B^\alpha_\perp \delta_{\omega,\cP_+ } B^\perp_\alpha \big] \,.
\end{equation}
The NRQCD fields $\psi_{\textbf{p}}$ and $\chi^\dagger_{- \textbf{p}}$ annihilate the heavy quark and antiquark fields, respectively. 
From now on we will drop the $({\bf 1},{}^3S_1)$ label. Note we correct a typo in Ref.~\cite{Fleming:2002rv} where the Kronecker delta has the incorrect label operator. 
The Upsilon contains no collinear quanta, so the current factors into an usoft piece containing the heavy quark spinors and a collinear piece containing the trace over the SCET gluon fields. The usoft fields can not ``talk" to the collinear fields due to color-transparency. We have simplified the above expression 
by fixing the momenta to be those of the the particular decay we are interested in. Strictly speaking this
can only be done after taking the matrix element of the operator above between external states, which  is given by
\begin{eqnarray} \label{amp1}
\langle J^\mu \rangle &=& \langle X_u | \chi^\dagger_{- \textbf{p}}
\Lambda\!\cdot\! \mathbf{\sigma}^\mu  \psi_{\textbf{p}} | \Upsilon \rangle  
 \sum_\omega e^{-i(Mv +M n/2)\cdot z}
 \Gamma_g (\omega; \mu)
 \langle X_c | \textrm{Tr}\big[ B^\alpha_\perp \delta_{\omega,\cP_+} B^\perp_\alpha \big] | 0 \rangle
 \nn \\
 & & \Longrightarrow  \langle X_u | \chi^\dagger_{- \textbf{p}}
\Lambda\!\cdot\! \mathbf{\sigma}^\mu  \psi_{\textbf{p}} | \Upsilon \rangle  
 \int^1_{-1} dx \, \Gamma_g (x ; \mu) \phi_g(x ; \mu ) \, .
 \end{eqnarray}
The outgoing state, $X_c$, is a jet with total momentum
$p^- = M_{\Upsilon}$, fixed by the mass of the decaying Upsilon. The kinematics are similar to the
meson light-cone wave function, and we therefore expect the running of the collinear matrix element 
$\phi_g(\omega ; \mu )$ which appears here in Upsilon decay to be the same as the running of the light-cone wave-function of a meson. Note the usoft matrix element does not run \cite{Hautmann:2001yz}.

\section{Running of Operators}

In SCET large logarithms are summed using the renormalization group equations (RGEs). In the case
we are interested in there are two LCAs,  the matrix elements of the operators in Eq.~(\ref{leadingops}), and they mix with each other. This will 
result in a coupled differential equation. In addition the LCAs under consideration is 
functions of the momentum fraction $x$, which makes the RGE
an integro-differential equation. 

The bare SCET operators, denoted by a zero superscript, are related to the renormalized operators
through a counterterm:
\begin{equation}\label{renorm}
O^{(0)}_a(x) = \int  d y  \, Z_{ab}(x,y ; \mu) O_b(y ; \mu)
\hspace{10ex} a,b = g,  q\,,
\end{equation}
where the bare  operator does not depend on  the scale $\mu$. Differentiating this equation with respect to $\mu$ and using the identity
\begin{equation}\label{id}
\int d z  \, Z_{ab}(z,x ; \mu)  Z^{-1}_{ca}(z, y ; \mu) 
= \delta_{bc} \, \delta(x - y ) \,.
\end{equation}
we obtain
\bea\label{anomdim}
\int  d y  \, 
Z_{ab}(x,y ; \mu) \, \mu \frac{d}{d \mu} O_b(y; \mu) & = &  
-  \int  d y  \,
\bigg( \mu \frac{d}{d \mu} Z_{ab}(x,y ; \mu) \bigg) 
O_b(y; \mu)
\nn \\
\Longrightarrow \mu \frac{d}{d \mu} O_c(z; \mu)  & = & 
-\int d  y \, O_b(y; \mu) \int d  x \,
Z^{-1}_{ca}(x, z ; \mu) 
\bigg( \mu \frac{d}{d \mu} Z_{ab}(x,y ; \mu) \bigg) 
\nn \\
 &=&  -\int dy \, 
\gamma_{cb}(z,y ; \mu) O_b(y ; \mu ) \,,
 \eea
with $\gamma_{cb}(z,y ; \mu)$ the anomalous dimension.  

The running of the short-distance coefficient is obtained as a consequence of the scale independence of the full theory current. For example differentiating both sides of Eq.~(\ref{matching}) with respect to $\mu$ gives zero on the left-hand side, which then gives a relationship between the running of the operators and the coefficient function. Generally the QCD current is matched onto the full set of operators given in Eq.~(\ref{leadingops}), and differentiating we obtain
\bea
 0 &=& \mu \frac{d }{d \mu} 
\int d x \, \Gamma_a (x; \mu) O_a(x ; \mu )
 \\
& = &  \int d x \,\left[  O_a(x ; \mu ) \mu \frac{d }{d \mu} 
\Gamma_a (x; \mu) +
  \Gamma_a (x; \mu)
\mu \frac{d }{d \mu}  O_a(x ; \mu )\right]
\nn \\
&=& \int d x \,\left[  O_a(x ; \mu ) \,\mu \frac{d }{d \mu} 
\Gamma_a (x; \mu) 
-    \int d y \,  \Gamma_a (x; \mu)
\gamma_{ab}(x, y ; \mu) O_b(y ; \mu)\right]
\nn \\
&=& \int d x \,  O_a(x ; \mu )  \left[ \mu \frac{d }{d \mu} 
\Gamma_a (x; \mu) 
 -     \int d y  \,  \Gamma_b (y; \mu)
\gamma_{ba}(y, x ; \mu) \right] = 0 \,, \nn
\eea
where we use the result of Eq.~(\ref{anomdim}) in obtaining the penultimate expression above, and interchanged the $x$ and $y$ (and $a$ and $b$ labels) in the second term to obtain the final expression. Since this must hold for any value of $x$, this equation implies an RGE for the coefficient function
 \begin{equation}\label{RGEcoeff}
 \mu \frac{d }{d \mu} \Gamma_a (x; \mu)  =     
 \int d y \,  \Gamma_b (y; \mu) \gamma_{ba}(y, x ; \mu) \,.
\end{equation}

The renormalization $Z$ can be calculated in perturbation theory. In dimensional 
regularization ($\overline{MS}$ scheme)  we obtain to ${\cal O}(\alpha_s)$:
\begin{equation}\label{zexp}
Z_{ab}(x, y ; \mu) = \delta_{ab} \, \delta(x - y )
+ \frac{1}{\epsilon} \frac{\alpha_s(\mu)}{2\pi} P_{ab}(x, y ) \,.
\end{equation}
In order to satisfy Eq.~(\ref{id}) we must have
\begin{equation}\label{invzexp}
Z^{-1}_{ab}(x, y ; \mu) = \delta_{ab} \, \delta(x - y )
- \frac{1}{\epsilon} \frac{\alpha_s(\mu)}{2\pi} P_{ab}(y,x ) \,,
\end{equation}
where on the right hand side of the equation above the $x$ and $y$ have been reversed. We now proceed to calculate $Z$ for the matrix element  which arises in Upsilon decay.

The Feynman diagrams which are needed to calculate $Z$ are shown in Figure~\ref{feyndiag}, while the Feynman rules for the operator vertices are given in Appendix \ref{appRules}. We show only those diagrams that are non-zero in dimensional regularization where the infrared is 
regulated by choosing $p_1^2=p_2^2 = 0$ and $p_1\cdot p_2 \neq 0$, with $p_{1,2}$ the momenta of the final state particles.
\begin{figure}[t]
\centerline{\includegraphics[width =6in]{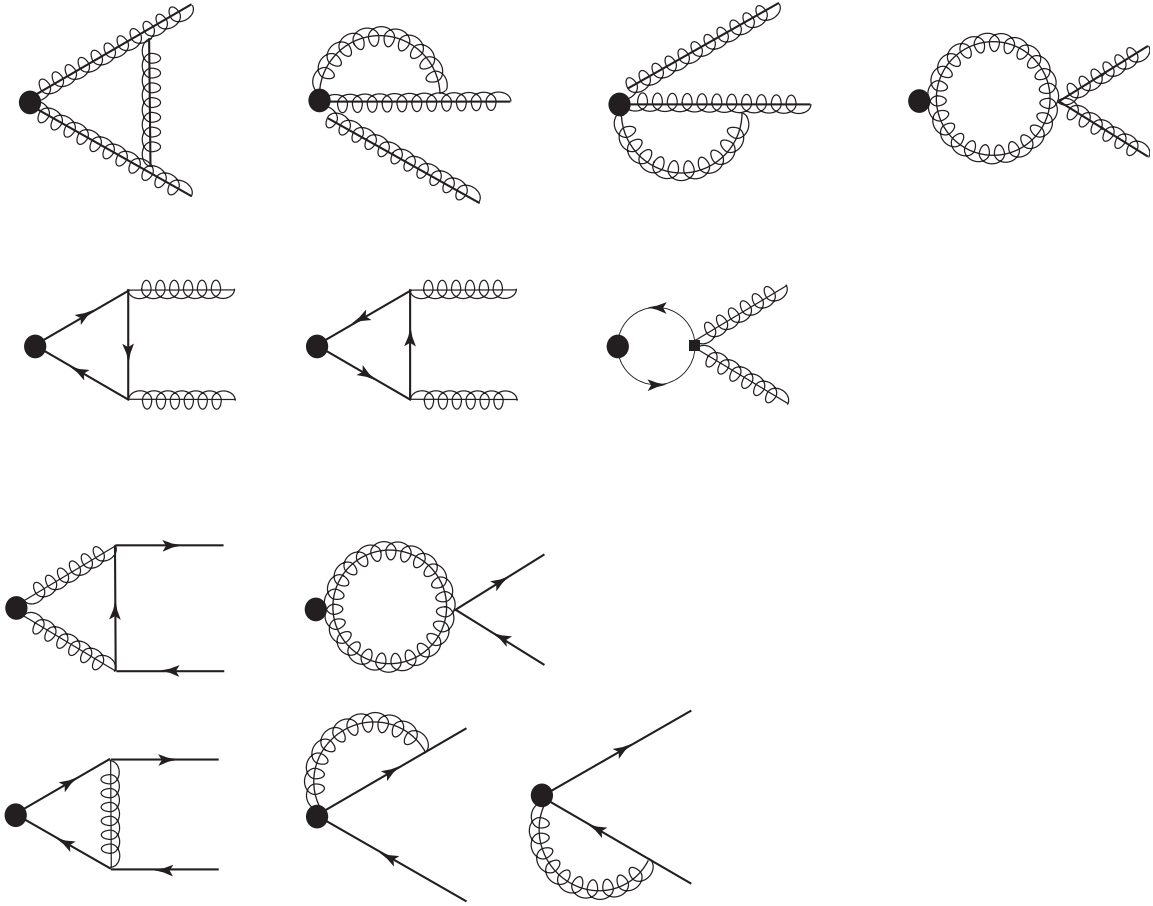}}
\caption{\it  One loop renormalization: a) glue to glue, b) quark to glue, c) glue to quark, 
d) quark  to quark.  The quark and gluon lines all represent collinear particles.
\label{feyndiag}}
\end{figure}
The divergent piece of the amplitude for each set is
\bea
{\cal M}_{\rm div} &=& \frac{\alpha_s}{2\pi\epsilon} \int  d y  \, \Delta_{ab} (x, y ) \Gamma_a(x;\mu) \phi_b(y; \mu)
\hspace{10ex} i,j = q,g \hspace{1ex} \,,
 \\
\Delta_{gg} (x, y ) &=& -C_A\,
\bigg\{ \bigg[ \frac{x^2 +  y^2}{(1 + x)(1-y)} +\frac{1}{2}
- \frac{1}{(x-y )_+} \bigg] \Theta(x-y) + 
{}^{x \to -x}_{y \to  - y}  - \delta(x-y) \bigg\},
\nn \\
\Delta_{gq} (x, y ) &=& \frac{C_F}{2}
\bigg\{  \frac{1-x+ 2 y}{(1-x)(1+y)} \Theta(y-x)
-  {}^{x \to -x}_{y \to  - y} \bigg\},
\nn \\
\Delta_{qg} (x, y ) &=& \frac{n_f}{2}\left\{
\frac{1 - x}{1-y}(1 - y + 2 x)\Theta(x-y) - 
{}^{x \to -x}_{y \to  - y}  \right\},
\nn \\
\Delta_{qq} (x, y ) &=& C_F\,
\bigg\{ \bigg[ \frac{1-x}{1-y}\bigg(\frac{1}{2} + \frac{1}{x-y}\bigg)_+
\Theta(x-y) +  {}^{x \to -x}_{y \to  - y}  \bigg]
+ \frac32  \delta(x-y) \bigg\} \,.
\nn
\eea
In obtaining these expression we made use of the property $\Gamma_a(-x)=\Gamma_a(x)$, which
is a consequence of the invariance of the product of operator and coefficient function under charge conjugation. These divergent amplitudes are canceled by the renormalization $Z_{ab}$. First we invert Eq.~(\ref{renorm}) and take the matrix element 
\bea
\phi_a(x) & = & \int  dy \,  \langle O^{(0)}_b(y) \rangle Z^{-1}_{ab}(y,x)
\nn \\
&=&  \int  dy \, \phi_b(y) Z_f Z^{-1}_{ab}(y,x) \,,
\eea
where $Z_f$ is the renormalization factor for the fields in the operator $O_a$. Expanding this to first order in dimensional regularization where $Z_f = 1+ \alpha_s(\mu) \delta_f / (2 \pi \epsilon)$, and using the expression in Eq.~(\ref{invzexp}) we obtain the equation which fixes the $P_{ab}$:
\begin{equation}
P_{ab}(x,y) =  \Delta_{ab} (x, y )+ \delta_f \delta_{ab} \delta(x-y) \,.
\end{equation}
From this we calculate the one loop expression for the anomalous dimension 
\begin{equation}
\gamma_{ab}(x,y;\mu) = -\frac{\alpha_s(\mu)}{\pi} P_{ab}(x,y)
\end{equation}
and substitute it into  Eq.~(\ref{RGEcoeff}) to obtain the one loop RGE
\bea
\mu \frac{d }{d \mu} \Gamma_a(x;\mu) & = &
-\frac{\alpha_s(\mu)}{\pi} \int dy \, \Gamma_b(y;\mu) P_{ba}(y,x)
\nn \\
&=& 
-\frac{\alpha_s(\mu)}{\pi} \int dy \, \Gamma_b(y;\mu) \bigg[   \Delta_{ba} (y, x )+ \delta_f \delta_{ab} \delta(x-y) \bigg]
\eea

With this result in hand we can solve the RGE by diagonalizing. The first step is to expand the coefficient funtions in a basis which is diagonal under the convolution with the $P_{ab}$. This basis is provided by the Gegenbauer polynomials~\cite{Lepage:1979zb,Efremov:1978rn,Chase:hj}:
\bea
\Gamma_{q} (x,\mu) &=& \sum_{n \rm{\ odd}} a^{(n)}_{q}(\mu) C^{3/2}_n(x)   \,,
\\
\Gamma_{g} (x,\mu) &=& \sum_{n \rm{\ odd}} a^{(n)}_{g}(\mu) (1-x^2)C^{5/2}_{n-1}(x)\,,
\nn
\eea
where the restriction to odd $n$ is required by Bose symmetry for the gluons.
Substituting these expansions into the evolution equations yields coupled ordinary differential 
equations
\begin{equation}\label{rge1}
\mu \frac{d}{d\mu} 
\left( \begin{array}{c}
a^{(n)}_{q} \\
a^{(n)}_{g}
\end{array} \right) 
 = -\frac{\alpha_s(\mu)}{\pi}
\left( \begin{array}{cc}
\gamma^{(n)}_{q\bar{q}} & \gamma^{(n)}_{gq} \\
\gamma^{(n)}_{qg} & \gamma^{(n)}_{gg} 
\end{array} \right)
\left( \begin{array}{c}
a^{(n)}_{q} \\
a^{(n)}_{g}
\end{array} \right)  \,,
\end{equation}
where 
\bea
\gamma^{(n)}_{q\bar{q}} &=&
C_F \bigg[ \frac{1}{(n+1)(n+2)} -\frac{1}{2} - 2 \sum^{n+1}_{i=2} \frac{1}{i} \bigg]\,,
\\
\gamma^{(n)}_{gq} &=&
\frac{1}{3}C_F \frac{n^2 + 3n +4}{(n+1)(n+2)}\,,
\nn \\
\gamma^{(n)}_{qg} &=&
3 n_f \frac{n^2 + 3n +4}{n(n+1)(n+2)(n+3)}\,,
\nn \\
 \gamma^{(n)}_{gg} &=&
 C_A \bigg[ \frac{2}{n(n+1)} + \frac{2}{(n+2)(n+3)}- \frac{1}{6} - 
 2 \sum^{n+1}_{i = 2} \frac{1}{i} \bigg] - \frac{1}{3}n_f\,.
 \nn
 \eea
The RGE in Eq.~(\ref{rge1}) can be diagonalized through a simlarity transformation 
resulting in 
\begin{equation}
\mu\frac{d}{d \mu} \mathbf {a}^{(n)} = -\frac{\alpha_s(\mu)}{\pi}
\mathbf{\Lambda}  \mathbf {a}^{(n)} \,,
\end{equation}
where the matrix $\mathbf{\Lambda} $ is diagonal and has eigenvalues
\begin{equation}
\lambda^{(n)}_\pm = \frac{1}{2} \big[ \gamma^{(n)}_{gg} +  \gamma^{(n)}_{q\bar{q}} 
\pm \Delta \big] \,,
\hspace{2.5ex} \mbox{with} \hspace{2.5ex} 
\Delta = 
\sqrt{ (\gamma^{(n)}_{gg} -  \gamma^{(n)}_{q\bar{q}})^2 + 
                 4 \gamma^{(n)}_{gq} \gamma^{(n)}_{qg} } \,.
\end{equation}
The eigenvector $ \mathbf {a}^{(n)} $ is
\begin{equation}\label{avec}
\mathbf {a}^{(n)} = 
\left( \begin{array}{c}
a^{(n)}_+ \\
a^{(n)}_-
\end{array} \right) = 
\left( \begin{array}{c}
a^{(n)}_q \gamma^{(n)}_{qg}  - 
a^{(n)}_g ( \lambda^{(n)}_- - \gamma^{(n)}_{gg} ) \\
a^{(n)}_q \gamma^{(n)}_{qg}  -
a^{(n)}_g (\lambda^{(n)}_+ -\gamma^{(n)}_{gg} )
\end{array} \right)\,.
\end{equation}
The diagonalized RGE is simple to solve, giving
\begin{equation}
a^{(n)}_\pm(\mu) = \bigg[ \frac{\alpha_s(M)}{\alpha_s(\mu)} \bigg]^{-2 \lambda_\pm / \beta_0}
a^{(n)}_\pm(M) \,,
\end{equation}
where $\beta_0 = 11 - 2 n_f / 3$. 
The equations above can be inverted to obtain
\bea\label{invertavec}
a^{(n)}_g(\mu) &=& \frac1\Delta\left(a^{(n)}_+(\mu) - a^{(n)}_-(\mu)\right)\,,
\\
a^{(n)}_q (\mu) &=& 
a^{(n)}_+(\mu) \frac{ \lambda^{(n)}_+ - \gamma^{(n)}_{gg} }{\Delta\gamma^{(n)}_{qg}}  + 
a^{(n)}_-(\mu) \frac{ \gamma^{(n)}_{gg} - \lambda^{(n)}_- }{\Delta\gamma^{(n)}_{qg}}
\,.
\nn
\eea
We can now include the running of the coefficients to get a result for the resummed gluon 
and quark coefficient:
\bea\label{runcoeff}
a^{(n)}_g(\mu) &=& \frac1\Delta
a^{(n)}_+(M) \bigg[ \frac{\alpha_s(M)}{\alpha_s(\mu)} \bigg]^{-2 \lambda^{(n)}_+ / \beta_0}  - 
\frac1\Delta
a^{(n)}_-(M) \bigg[ \frac{\alpha_s(M)}{\alpha_s(\mu)} \bigg]^{-2 \lambda^{(n)}_- / \beta_0} \,,
\\
a^{(n)}_q(\mu)  &=& 
a^{(n)}_+(M) \frac{ \lambda^{(n)}_+ - \gamma^{(n)}_{gg} }{\Delta\gamma^{(n)}_{qg}}
\bigg[ \frac{\alpha_s(M)}{\alpha_s(\mu)} \bigg]^{-2 \lambda^{(n)}_+ / \beta_0}  + 
a^{(n)}_-(M) \frac{ \gamma^{(n)}_{gg} - \lambda^{(n)}_- }{\Delta\gamma^{(n)}_{qg}}
\bigg[ \frac{\alpha_s(M)}{\alpha_s(\mu)} \bigg]^{-2 \lambda^{(n)}_- / \beta_0} \,.
\nn
\eea

So far our results have been general, and can be used for not only Upsilon decay, but any process with helicity-zero, flavor- and color-singlet wavefunctions.  The process dependence will come in the boundary conditions.
For Upsilon decay, the matching coefficient for the 
quark operator is zero at leading order, while the matching coefficient for the gluon operator is 
a constant $\kappa$.  We will normalize our matrix element so that $\kappa = 1$. Having expanded the matching coefficients in Gegenbauer polynomials we  
determine
\bea
a^{(n)}_q(M) &=& 0\,,
\\
a^{(n)}_g(M) &=& \frac{4}{3 f_{5/2}^{(n)}} \,,
\nn
\eea
where 
\begin{equation}
f_{5/2}^{(n)} = \frac{n(n+1)(n+2)(n+3)}{9(n+3/2)}
\end{equation}
is the normalization of $C_{n-1}^{5/2}(x)$.

Using the relations in Eq.~(\ref{invertavec}) we determine the initial conditions for the 
componenets of $\mathbf{a}$:
\bea
a^{(n)}_+(M) &=& (\gamma^{(n)}_{gg} - \lambda^{(n)}_-) \,a_g^{(n)}(M) \,,
\\
a^{(n)}_-(M) &=& (\gamma^{(n)}_{gg} - \lambda^{(n)}_+) \,a_g^{(n)}(M) \,.
\nn
\eea
These can be substituted into Eq.~(\ref{runcoeff}) to obtain the final result:
\bea\label{resco}
a^{(n)}_q(\mu) &=& 
\frac{\gamma^{(n)}_{gq}}{\Delta}\left\{ \bigg[ \frac{\alpha_s(M)}{\alpha_s(\mu)} \bigg]^{-2 \lambda^{(n)}_+ / \beta_0}  - \bigg[ \frac{\alpha_s(M)}{\alpha_s(\mu)} \bigg]^{-2 \lambda^{(n)}_- / \beta_0} \right\}a_g^{(n)}(M)\,,
\\
a^{(n)}_g(\mu)  &=& 
\left\{ \gamma_+^{(n)} \bigg[ \frac{\alpha_s(M)}{\alpha_s(\mu)} \bigg]^{-2 \lambda^{(n)}_+ / \beta_0} - 
\gamma_-^{(n)}\bigg[ \frac{\alpha_s(M)}{\alpha_s(\mu)} \bigg]^{-2 \lambda^{(n)}_- / \beta_0}\right\}a_g^{(n)}(M)\,,
\eea
where
\bea
\gamma_\pm^{(n)} = \frac{\gamma_{gg}^{(n)} - \lambda^{(n)}_\mp}{\Delta}.
\eea

\section{Resummed Rate}

\begin{figure}[t]
\centerline{ \includegraphics[width=2.5in]{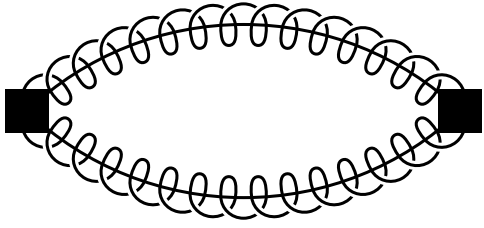}}
\caption{\it Feynman diagram for the leading order gluon jet function.
\label{gluevacloop}}
\end{figure}

\begin{figure}[t]
\centerline{ \includegraphics[width=2.5in]{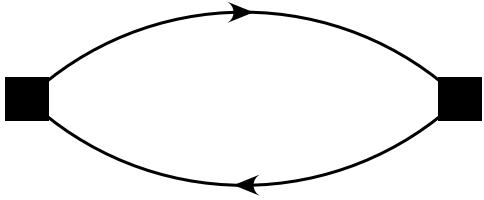}}
\caption{\it Feynman diagram for the leading order quark jet function.
\label{quarkvacloop}}
\end{figure}

The decay rate is proportional to the imaginary part of the forward scattering amplitude $T$. The expression for this ampitude was derived and given in Eq.~(59) of Ref.~\cite{Fleming:2002rv}\footnote{Here we fix a typo in that equation.},
\begin{equation}\label{TOP}
\textrm{Im} T(z) = \int dx \,  \frac{2M_\Upsilon}{M^2} H(x)  \int d\ell^+ S(\ell^+ , \mu ) 
\textrm{Im} J[Mx,\ell^+ + M(1-z);\mu]\,,
\end{equation}
where $H$ is a hard coefficient, $S(l^+)$ is the color-singlet shape function \cite{Rothstein:1997ac},
\bea
S(l^+) &=& \int\frac{dx^-}{4\pi} e^{-i l^+ x^-/2} \langle \Upsilon|T[\psi^\dagger_{- \textbf{p}} \mathbf{\sigma}_i  \chi_{-\textbf{p}} ](x^-) [\chi^\dagger_{- \textbf{p}'}
\mathbf{\sigma}_i \psi_{\textbf{p}'}](0) | \Upsilon \rangle  \nn\\
&=& \langle \Upsilon| \psi^\dagger_{- \textbf{p}} \mathbf{\sigma}_i  \chi_{-\textbf{p}} \delta(in\cdot\partial - l^+)  \chi^\dagger_{- \textbf{p}'} \mathbf{\sigma}_i \psi_{\textbf{p}'} | \Upsilon \rangle ,
\eea
and $\textrm{Im} J[Mx(\ell^+ + M(1-z));\mu]$ is the imaginary part of the jet function,
\bea
\label{oldjet}
\langle 0| T \textrm{Tr}
\big[B^\alpha_\perp \delta(\omega-i \bar{n}\!\cdot\!{\cal D}_+)B^\perp_\alpha\big](y)
\big[B^\beta_\perp \delta(\omega'-i \bar{n}\!\cdot\!{\cal D}_+)B^\perp_\beta\big] |0 \rangle
\nn \\
&& \hspace{-25ex}
= 2 i \delta(\omega-\omega') \int\frac{d^4k}{(2\pi)^4} e^{-ik\cdot y} J(\omega, k^+;\mu)\,,
\eea
where the labels $\omega$ and $\omega'$ are continuous and 
$i\bar{n}\!\cdot\!{\cal D}= \bnP+i\bar{n}\!\cdot\!\partial$ as discussed in Appendix~\ref{appRPI}.
Since Ref.~\cite{Fleming:2002rv} did not consider mixing, this was the only jet function.  We now generalize this to
\begin{equation}\label{newjet}
\langle 0| T [O_a(\omega; y) O_b(\omega' ; 0) ] | 0 \rangle = 2 i M\,\delta(\omega-\omega') \int\frac{d^4k}{(2\pi)^4} e^{-ik\cdot y} J_{ab}(\omega,k^+;\mu)\,,
\end{equation}
where $a,b = g,q$.  The hard coefficient gets modified to be
\begin{equation}
H_{ab}(x) = \frac{4}{3}\left(\frac{4g_s^2 e e_b}{3M}\right)^2\Gamma_a(x)\Gamma_b(x).
\end{equation}
If $a \neq b$, we get no contribution at this order in perturbation theory.  When $a = b = g$, we get exactly what was considered in Ref.~\cite{Fleming:2002rv}, pictured in Fig.~\ref{gluevacloop}.
The imaginary part of the jet function in this case is
\begin{equation}
{\rm Im } J_{gg}(\omega, k^+;\mu) = \frac{1}{8 \pi} \Theta( k^+).
\end{equation}
We now need the imaginary part of the quark jet function, picture in Fig.~\ref{quarkvacloop}.
The result is 
\begin{equation}
{\rm Im } J_{qq}(\omega , k^+;\mu) = \frac{N_c}{N_c^2-1} \frac{M^2 - \omega^2}{M^2}  \frac{1}{8 \pi} \Theta( k^+).
\end{equation}

Expanding the matching coefficient in Gegenbauer polynomials the integral over $\omega = M x$ in Eq.~(\ref{TOP}) may be be carried out, giving
\bea
\int^1_{-1}  dx  \, \Gamma_g^2(x,\mu) & = & \sum_{n  \rm{\ odd}}
\big[ a_g^{(n)}(\mu) \big]^2 \int^1_{-1} dx  \, C^{5/2}_{n-1}(x) C^{5/2}_{n-1}(x) (1-x^2)^2 
\\
&=& \frac{16}{9} \sum_{n \rm{\ odd}} \frac{1}{f_{5/2}^{(n)}}
\left[ \gamma_+^{(n)}
r(\mu)^{2 \lambda^{(n)}_+ / \beta_0}  - 
\gamma_-^{(n)}
r(\mu)^{2 \lambda^{(n)}_- / \beta_0} \right]^2
 \,, \nn
\eea
for the gluon jet function, and
\bea
\int^1_{-1}  dx  \,(1 - x^2) \Gamma_q^2(x,\mu) & = & \sum_{n  \rm{\ odd}}
\big[ a_q^{(n)}(\mu) \big]^2 \int^1_{-1} dx  \, (1-x^2) C^{3/2}_n(x) C^{3/2}_n(x) 
\\
&=& \frac{16}{9} \sum_{n \rm{\ odd}} \frac{f_{3/2}^{(n)}}{[f_{5/2}^{(n)}]^2} \frac{{\gamma^{(n)}_{gq}}^2}{\Delta^2}
\left[ r(\mu)^{2 \lambda^{(n)}_+ / \beta_0}  -  r(\mu)^{2 \lambda^{(n)}_- / \beta_0} \right]^2
 \,, \nn
\eea
for the quark jet function, where we have defined
\begin{equation}
r(\mu) = \frac{\alpha_s(\mu)}{\alpha_s(M)}\,,
\end{equation}
and
\begin{equation}
f_{3/2}^{(n)} = \frac{(n+1)(n+2)}{n+3/2}
\end{equation}
is the normalization of $C^{3/2}_n(x)$.  Using the results of Ref.~\cite{Fleming:2002rv}, the differential decay rate is
\bea\label{diffrate}
\frac{1}{\Gamma_0}\frac{d\Gamma_{\rm resum}}{dz} &=& \Theta(M_\Upsilon - M z) \frac{8z}9 
\sum_{n \rm{\ odd}} \left\{\frac{1}{f_{5/2}^{(n)}}
\left[ \gamma_+^{(n)}
r(\mu_c)^{2 \lambda^{(n)}_+ / \beta_0}  - 
\gamma_-^{(n)}
r(\mu_c)^{2 \lambda^{(n)}_- / \beta_0} \right]^2\right.
\\
&&\phantom{\Theta(M_\Upsilon - M z) \frac{8z}9 
\sum_{n \rm{\ odd}} []}\left.+
\frac{3 f_{3/2}^{(n)}}{8[f_{5/2}^{(n)}]^2}\frac{{\gamma^{(n)}_{gq}}^2}{\Delta^2}
\left[ r(\mu_c)^{2 \lambda^{(n)}_+ / \beta_0}  -  
r(\mu_c)^{2 \lambda^{(n)}_- / \beta_0} \right]^2\right\}\,,\nn
\eea
where $\mu_c = M\sqrt{1-z}$ is the collinear scale.
This results differs from the result in Ref.~\cite{Photiadis:1985hn}.  We agree with Ref.~\cite{Photiadis:1985hn} up to Eq.~(5) of that paper, up to some typos. However, following the method outlined in that paper, we still arrive at the first line of Eq.~(\ref{diffrate}).  Both Eq.~(\ref{diffrate}) and the result in Ref.~\cite{Photiadis:1985hn} reduce to the rate calculated in Ref.~\cite{Fleming:2002rv}, when the mixing is turned off.  The second line of Eq.~(\ref{diffrate}) comes from the quark jet function.  This adds a small contribution to the total resummed rate.  

While the logarithms that are summed in Eq.~(\ref{diffrate}) are important at large $z$, this formula should not be trusted away from the endpoint.  In order to match our resummed result onto the leading order result, we will interpolate between the two using the formula,
\begin{equation}\label{fulleq}
\frac{1}{\Gamma_0} \frac{d \Gamma_{\rm int}}{d z}= 
\bigg( \frac{1}{\Gamma_0} \frac{d \Gamma_{\rm LO}^{\rm dir}}{d z} - z \bigg)
+ \frac{1}{\Gamma_0} \frac{d \Gamma_{\rm resum}}{d z} \,,
\end{equation}
where \cite{firstRad}
\begin{equation}\label{LOrate}
\frac{1}{\Gamma_0}\frac{d\Gamma_{\rm LO\ direct}}{dz} = \frac{2-z}{z} + \frac{z(1-z)}{(2-z)^2} + 2\frac{1-z}{z^2}\ln(1-z) - 2\frac{(1-z)^2}{(2-z)^3}\ln(1-z)\,,
\end{equation}
and
\begin{equation}
\Gamma_0 = \frac{32}{27} \alpha \alpha_s^2 e_b^2\frac{\langle\Upsilon|\psi^\dagger_{- \textbf{p}} \mathbf{\sigma}_i  \chi_{-\textbf{p}}   \chi^\dagger_{- \textbf{p}'} \mathbf{\sigma}_i \psi_{\textbf{p}'} | \Upsilon \rangle}{m_b^2}\,.
\end{equation}
The term
in brackets in Eq.~(\ref{fulleq}) vanishes as $z \to 1$, leaving only the
resummed contribution in that region. Away from the endpoint the
resummed contribution combines with the $-z$ to give higher order in
$\alpha_s(M)$ corrections.  This is clear from Eq.~(\ref{diffrate}).  There are important corrections to this result due to fragmentation at low $z$ \cite{Catani:1995iz}.  However, since we are interested in the large $z$ region, we will neglect them in the following.  There may also be large color-octet corrections to the rate in the endpoint region \cite{Maltoni:1999nh,Bauer:2001rh,GarciaiTormo:2004jw}, which we will also neglect for the now.  We also compare our result to the resummed result where the mixing has been turned off, using
\begin{equation}\label{nomixrate}
\frac{1}{\Gamma_0}\frac{d\Gamma_{\rm no\ mix}}{dz} = \Theta(M_\Upsilon - M z) \frac{8z}9 
\sum_{n \rm{\ odd}} \frac{1}{f_{5/2}^{(n)}}
\bigg[ \frac{\alpha_s(\mu_c)}{\alpha_s(M)} \bigg]^{4 \gamma_{gg}^{(n)} / \beta_0}\,,
\end{equation}
in place of the full resummed result in Eq.~(\ref{fulleq}).

\begin{figure}[t]
\centerline{\includegraphics[width=5in]{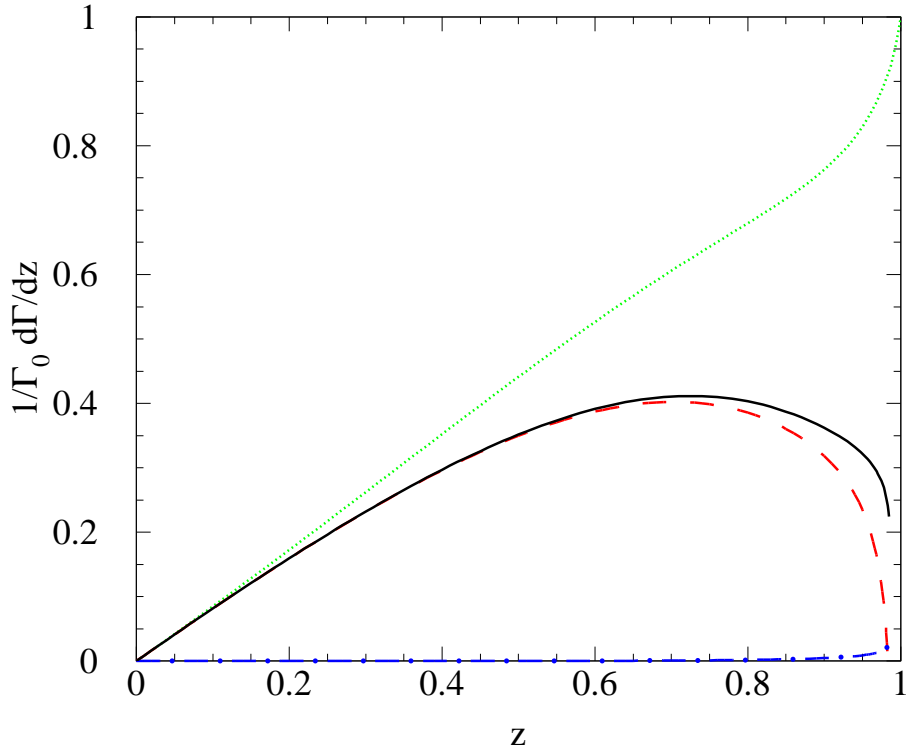}}
\caption{\it The color-singlet rate.  The dotted curve is the
tree-level direct rate.  The solid curve is the interpolated resummed direct
rate.  The dashed curve is the resummed rate with the mixing turned off.}
\label{mixfig}
\end{figure}
In Fig.~\ref{mixfig} we show the color-singlet, interpolated resummed rate, Eq.~(\ref{fulleq}) show as the solid line, compared to the leading tree level color-singlet result, Eq.~(\ref{LOrate}), shown as the dotted line.  As can be seen, the 
resummed rate turns over and decreases near the endpoint.  
Also shown in Fig.~\ref{mixfig} is the interpolated resummed rate with the mixing turned off, Eq.~(\ref{nomixrate}).  This is the same as the results of Ref.~\cite{Fleming:2002rv}.  The result without mixing is a fairly good approximation to the full result.  We also show the contribution coming from the quark jet alone as the dot-dashed line.

\section{Conclusions}

Radiative Upsilon decay at maximal photon energy is characterized by a photon recoiling against 
a jet of collinear particles.  Thus SCET is the appropriate effective field theory to study this 
kinematic situation.  The lowest order color-singlet QCD diagram for this process has the Upsilon decaying to a photon and a pair of gluons.  In a previous pair of papers \cite{Fleming:2002rv}, we
used SCET to investigate the endpoint behavior, summing kinematic logarithms.  However, we
neglected the possible mixing of the gluon pair with a quark--antiquark pair.  The full calculation, 
including the operator mixing, had been presented in the literature by Photiadis~\cite{Photiadis:1985hn}.  
As pointed out in Ref.~\cite{Photiadis:1985hn}, the radiative Upsilon decay at the endpoint has the
same evolution equations as the flavor-singlet light-cone wavefunction evolution.  

Therefore, we have calculated the flavor- and color-singlet, helicity-zero light-cone amplitude evolution using SCET, with the goal in mind of studying the photon endpoint spectrum in radiative Upsilon 
decay.  We find that SCET does reproduce the evolution equations for the light-cone amplitudes presented previously in the literature \cite{Chase:hj}.  When applying this to Upsilon decay, we 
however disagree with Ref.~\cite{Photiadis:1985hn}, although numerically the results are similar.  

With the inclusion of the operator mixing, we have a complete, leading logarithm result for the color-singlet contribution to radiative Upsilon decay at the endpoint.  Combining this with the leading logarithm result for the color-octet contribution at the endoint \cite{Bauer:2001rh,GarciaiTormo:2004jw}, 
and the photon fragmentation results at low $z$ \cite{Catani:1995iz,Maltoni:1999nh}, we can hope to obtain an accurate prediction for the photon spectrum over the full kinematic range.  

\acknowledgments 
This work was supported in part by the Department of Energy under grant number
DOE-ER-40682-143 and in part by the National Science 
Foundation under Grant No.~PHY-0244599

\appendix
\section{Operators of Continuous labels}\label{appRPI}

In this Appendix we explain the relationship of SCET operators defined using a discrete label to those defined using a continuous label. As a concrete example we consider the current in Eq.~(\ref{amp1}), which involves the gluon operator. The matrix element of the collinear operator in the first line is 
\begin{equation}\label{disc}
\langle X_c | \textrm{Tr} \big[ B^\alpha_\perp \delta_{\omega,\cP_+} B^\perp_\alpha \big] (x) | 0 \rangle
\,,
\end{equation}
where we have made explicit the space-time dependence. The expression above is defined for a 
discrete label $\omega$. However, we could write down an operator involving a continuous label $\omega_c$
\begin{equation}\label{cont}
\langle X_c | \textrm{Tr} \big[B^\alpha_\perp 
\delta(\omega_c-i\bar{n}\!\cdot\!{\cal D}_+ ) 
B^\perp_\alpha \big] (x) | 0 \rangle
\,,
\end{equation}
where $i\bar{n}\!\cdot\!{\cal D}_+ = \bnP_+ - i \bar{n}\!\cdot\!\overleftarrow{\partial} +  i \bar{n}\!\cdot\!\overrightarrow{\partial}$. Note that the sum over $\omega$ in Eq.~(\ref{amp1}) is now replaced with an integral over $\omega_c$. The delta function must be understood as 
\begin{equation}
 \delta(\omega_c-i\bar{n}\!\cdot\!{\cal D}_+ ) \equiv 
 \delta_{\omega,\bnP_+} \delta(k-i\bar{n}\!\cdot\!{\cal \partial}_+ ) 
 \end{equation}
 where $\omega_c = \omega + k$ with $\omega \sim M$ discrete, and $k \sim \LQCD$ continuous. The integral over $\omega_c$ must then be understood as a sum over $\omega$ and an integral over $k$. 
The expression in Eq.~(\ref{cont}) can be expanded in powers of 
$i \bar{n}\!\cdot\!\partial / \bnP \sim \LQCD/M$, where the leading term is just the operator in Eq.~({\ref{disc}). Thus the continuous operator is just the discrete operator plus high-order corrections. However, in an EFT it is only sensible to include higher order corrections in a leading order operator if all of the subleading terms run the same way as the leading term (i.e., they all have the same anomalous dimension). This is only true if there is a symmetry which enforces this condition. In this case the symmetry is a specific reparameterization invariance known as RPI (a)~\cite{Chay:2002vy,Manohar:2002fd}. In essence this RPI is the statement that there is no unique way to decompose the large label momentum and the continuous residual momentum. This implies that reparamterization invariant operators must be built out of $i\bar{n}\!\cdot\!{\cal D}$, and that such an operator runs in a specific way. As a result any subleading operators that are due to an expansion of  $i\bar{n}\!\cdot\!{\cal D}$ in powers of 
$i \bar{n}\!\cdot\!\partial / \bnP$ must have the same running.

\section{Feynman rules}\label{appRules}

In this Appendix we give the Feynman rules derived from the color-singlet operators given in Eq.~(\ref{leadingops}), which we repeat here:
\bea\label{leadingops2}
O_{g}(\omega_1,\omega_2) &=& \bnP
{\rm Tr}[B_{\perp \omega_1}^\alpha \, B_{ \perp \omega_2 }^\beta] g^\perp_{\alpha\beta} \,,
\nn\\
O_{q}(\omega_1,\omega_2) &=& \bar\chi_{n,\omega_1}\frac{\bnslash}{2}\chi_{n,\omega_2}\,.
\eea
The fields $B_{\perp,\omega}^\alpha$ and $\chi_{n,\omega}$ are built using the collinear Wilson line in order to obtain gauge invariant objects.  We thus have an infinite number of Feynman rules encoded in each operator of Eq.~(\ref{leadingops2}).  Here, we will give the corresponding Feynman rules necessary for calculating the anomalous dimension of the operators at one loop, namely the operators to order $g_s^0$ and $g_s^1$.  We will always define our momentum to be incoming.  The Feynman rules are shown in Fig.~\ref{FRfig}.
\begin{figure}[t]
\centerline{\hspace{2cm}\includegraphics[width=3in]{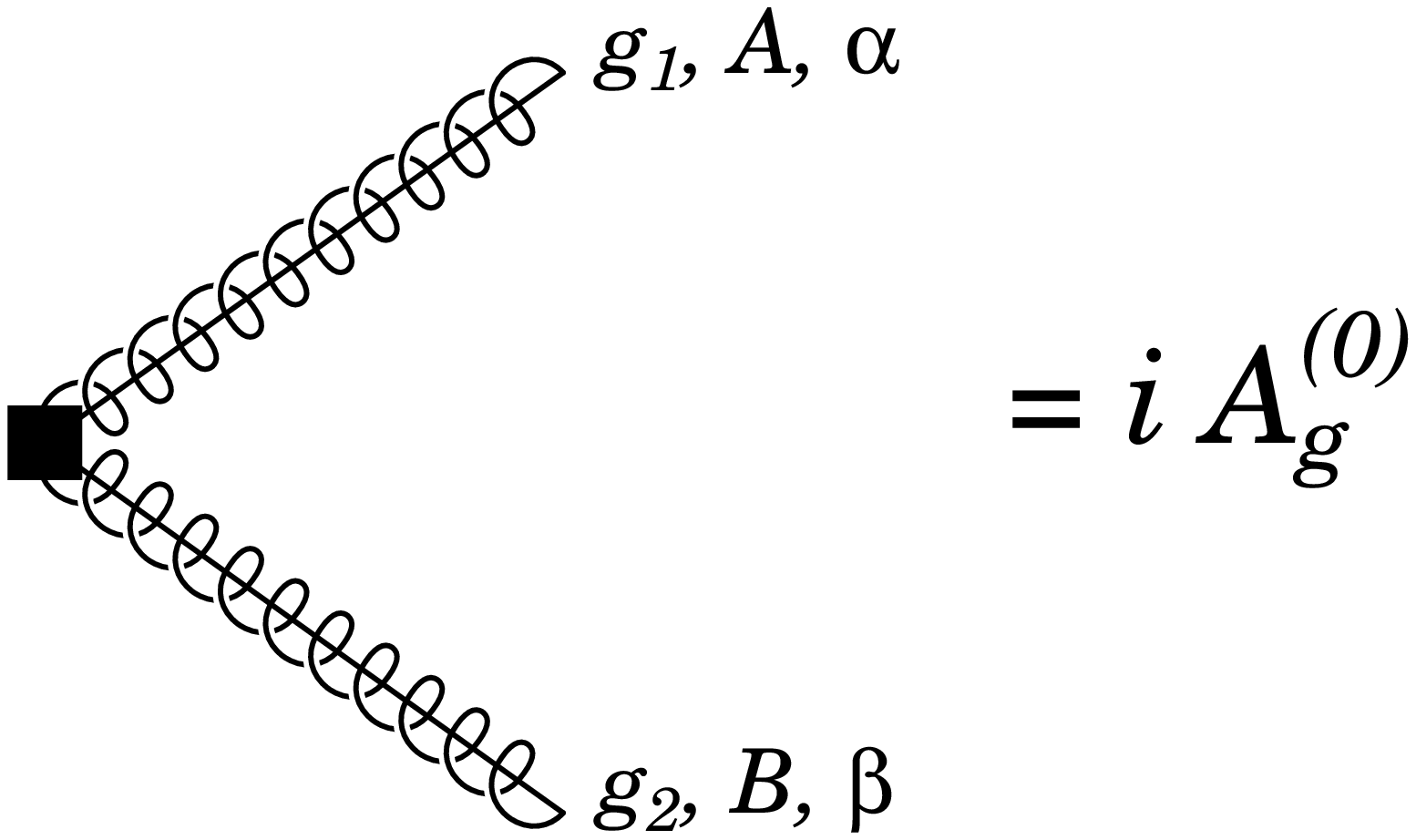}
\qquad
\includegraphics[width=3in]{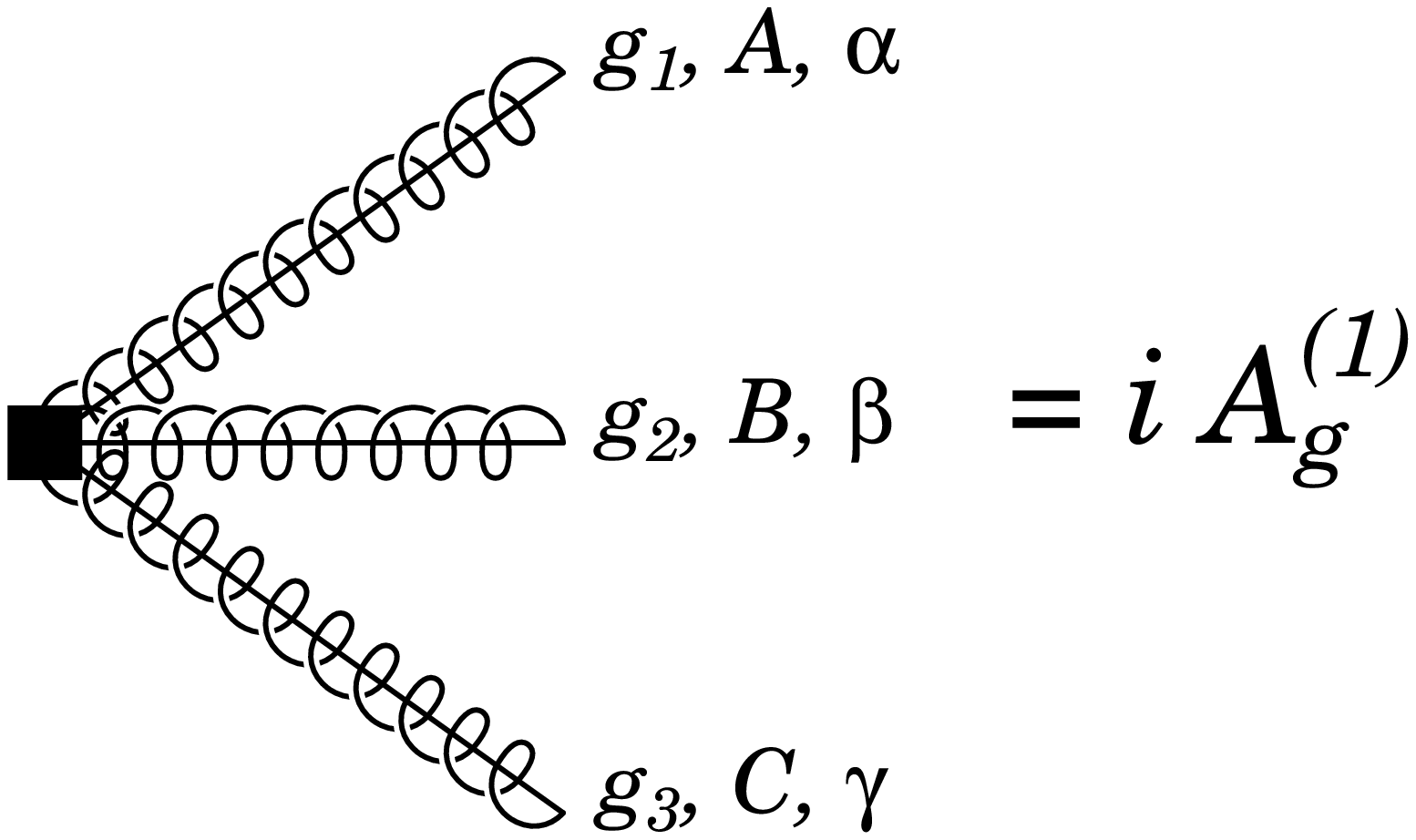}}
\vspace{1cm}
\centerline{\hspace{2cm}\includegraphics[width=3in]{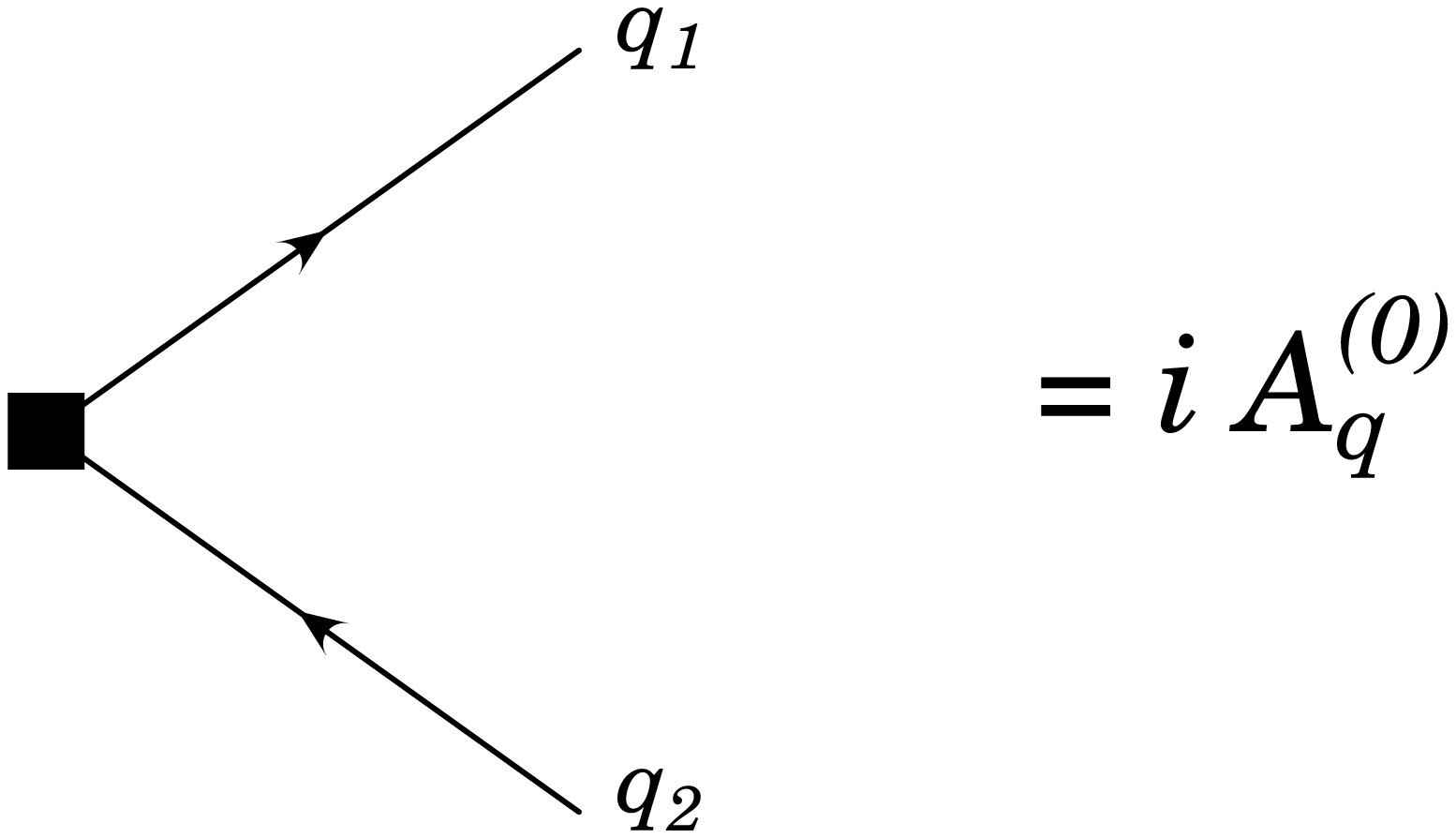}
\qquad
\includegraphics[width=3in]{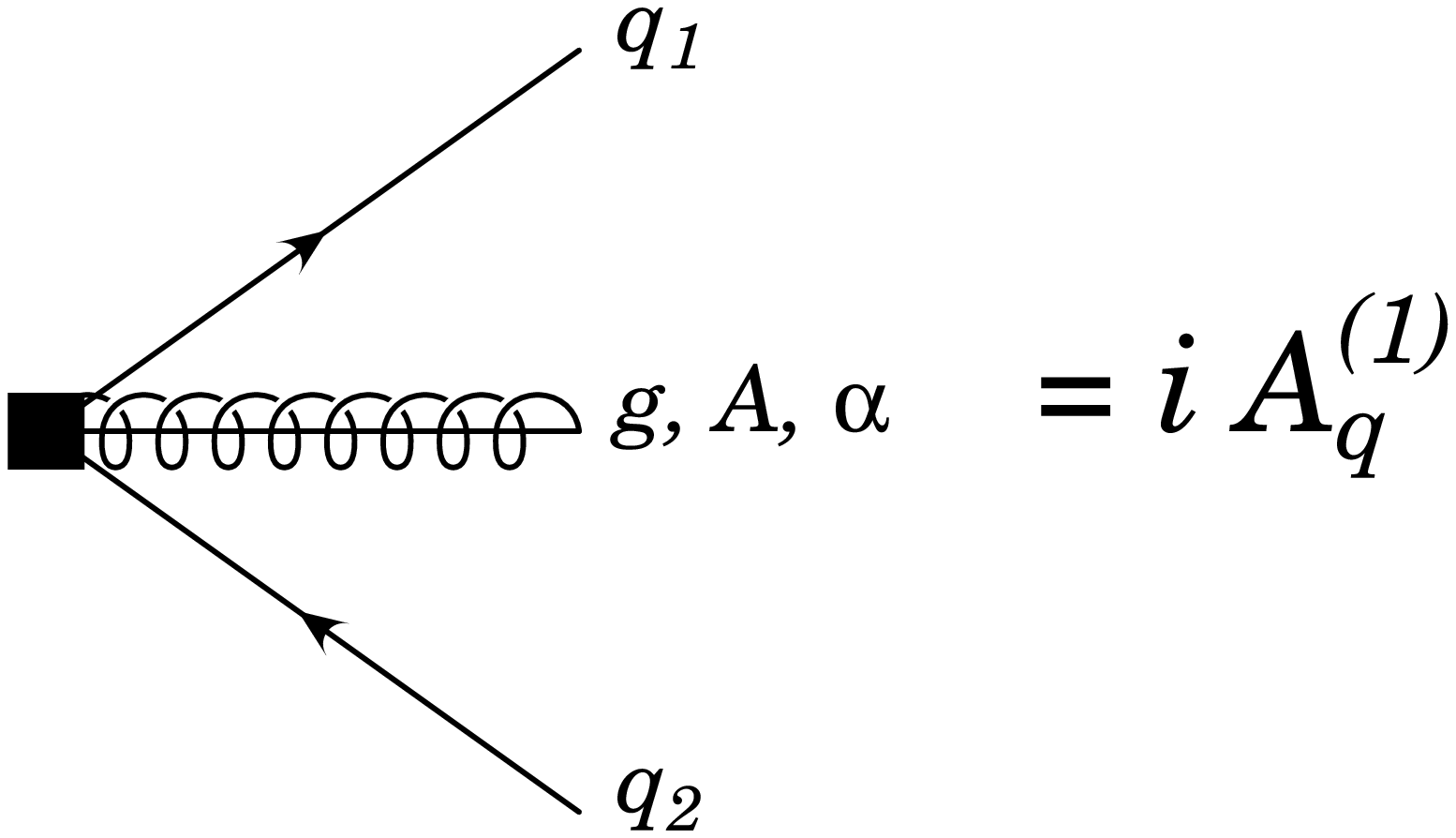}}
\caption{\it The Feynman rules corresponding to the color-singlet operators of Eq.~(\ref{leadingops2}).}
\label{FRfig}
\end{figure}

We begin with the gluon operator.  We have
\begin{equation}
O_{g}(\omega_1,\omega_2) = 2 \bnP
{\rm Tr}[B_{\perp}^\alpha\,\delta(\omega_+ - \bnP_+) \,\delta(\omega_-  - \bnP_-) \, 
 B_{ \perp}^\beta] g^\perp_{\alpha\beta}\,,
\end{equation}
where $\omega_\pm = \omega_1 \pm \omega_2$, $\bnP_\pm = \bnP \pm \bnP^\dagger$, and the factor of 2 will cancel the Jacobian from changing from $\omega_{1,2}$ to $\omega_\pm$.  The delta function $\delta(\omega_- - \bnP_-)$ will constrain the sum of the momenta to be the total energy of the jet, which in our case is $M_\Upsilon$.    We are therefore only interested in the Feynman rule for
\begin{equation}
\bar O_{g}(\omega_-) = \bnP {\rm Tr}[B_{\perp}^\alpha \,\delta(\omega_+  - \bnP_+) \, 
 B_{ \perp}^\beta] g^\perp_{\alpha\beta}\,.
\end{equation}
Expanding out the $B_\perp^\mu$ to leading order in $g_s$, we get the order $g_s^0$ Feynman rule
\bea
i{\cal A}_g^{(0)} &=& -\frac{i}{2}M\delta^{AB}\,\left[ \delta(\omega_+ + \bn\cdot g_1 - \bn\cdot g_2) + \delta(\omega_+ - \bn\cdot g_1 + \bn\cdot g_2)\right]\nn\\
&&\times
g_{\mu\nu}^\perp\left(g_\perp^{\alpha\mu} - \frac{g_{1\perp}^\mu \bn^\alpha}{\bn\cdot q_1}\right)
\left(g_\perp^{\beta\nu} - \frac{g_{2\perp}^\nu \bn^\beta}{\bn\cdot q_2}\right).
\eea
At order $g_s^1$ we get
\bea
i{\cal A}_g^{(1)} &=& \frac{g_s}{2} M f^{ABC} \,
\left[\delta(\omega + \bn\!\cdot\! g_1 -  \bn\!\cdot\! g_2 -  \bn\!\cdot\! g_3) 
 + \delta(\omega -  \bn\!\cdot\! g_1 +  \bn\!\cdot\! g_2 +  \bn\!\cdot\! g_3)\right]\,g^\perp_{\mu\nu}\nn\\ 
&&\quad\times
\left(g_\perp^{\alpha\mu} - \frac{g_{1\perp}^\mu \bn^\alpha}{\bn\!\cdot\! g_1}\right)
\left\{ 
\left(g_\perp^{\gamma\nu} - \frac{g_{3\perp}^\nu \bn^\gamma}{\bn\!\cdot\! (g_2+g_3)}\right)
\frac{\bn^\beta}{\bn\!\cdot\! g_2} -
\left(g_\perp^{\beta\nu} - \frac{g_{2\perp}^\nu \bn^\beta}{\bn\!\cdot\! (g_2+g_3)}\right)
\frac{\bn^\gamma}{\bn\!\cdot\! g_3}
\right\}
\nn\\
&& +\, [(1,\alpha) \to (2,\beta) \to (3,\gamma) \to (1,\alpha) ]  \nn\\
&& + \, [(1,\alpha) \to (3,\gamma)  \to (2,\beta)\to (1,\alpha) ] \,.
\eea

We can similarly rewrite our quark operator as
\bea
O_{q}(\omega_1,\omega_2) &=& 2\bar\xi_{n,p_1}\frac{\bnslash}{2}\,\delta(\omega_+ - \bnP_+) \,\delta(\omega_-  - \bnP_-) \, \xi_{n,p_2}\nn\\
\Longrightarrow \bar O_{q}(\omega_-) &=& \bar\xi_{n,p_1}\frac{\bnslash}{2} \,\delta(\omega_+  - \bnP_+) \, \xi_{n,p_2}\,.
\eea
This gives the order $g_s^0$ Feynman rule
\begin{equation}
i{\cal A}_q^{(0)} = i \bar\xi_{n} \frac{\bnslash}{2} \xi_{n}
\delta(\omega_+ + \bn\cdot q_1 - \bn\cdot q_2)\,,
\end{equation}
where again, the momentum is defined to be incoming.  The order $g_s^1$ Feynman rule is
\begin{equation}
i{\cal A}_q^{(1)} = i g_s\frac{\bn^\alpha}{\bn\cdot g} \bar\xi_{n} \frac{\bnslash}{2} T^A \xi_{n} 
\left[
\delta(\omega_+ + \bn\cdot q_1 - \bn \cdot q_2 - \bn \cdot g) -
 \delta(\omega_+ + \bn\cdot q_1 - \bn \cdot q_2 + \bn \cdot g)
\right]\,.
\end{equation}
%


\end{document}